\documentclass[a4paper,10pt,twocolumn]{article}

\usepackage{graphicx}
\usepackage{amssymb}
\usepackage{bm}
\usepackage{amsmath}
\usepackage{subfigure}

\title{Competing synapses with two timescales: a basis for learning and forgetting}

\author{Gaurang Mahajan\footnote{\texttt gaurang@bose.res.in} ~and Anita Mehta\footnote{\texttt anita@bose.res.in}\\
\\
S N Bose National Centre for Basic Sciences,\\
Block JD, Sector III, Salt lake, Kolkata 700 098, India \\
}

\begin{document}

\maketitle

\begin{abstract}
Competitive dynamics are thought to occur in many processes of learning involving synaptic plasticity. Here we show, in a game theory-inspired model of synaptic interactions, that the competition between synapses in their weak and strong states gives rise to a natural framework of learning, with the prediction of memory inherent in a timescale for `forgetting' a learned signal. Among our main results is the prediction that memory is optimized if the weak synapses are really weak, and the strong synapses are really strong. Our work admits of many extensions and possible experiments to test its validity, and in particular might complement an existing model of reaching, which has strong experimental support.
\end{abstract}

\section*{Introduction}

Multiple spatial and temporal scales occur in the dynamics of neural systems. 
Several recent studies~\cite{fusi05,smith06,juntani08,fusi07,cc08} have explored the consequences of having multiple built-in dynamical timescales in adaptation and learning mechanisms in particular, and its possible functional benefits. One such example is
the study of \cite{smith06}, which considered short-term motor learning and forgetting involved in the phenomenon of reaching. By means of numerical simulation, a two-state system containing a slow and a fast timescale was shown to provide a unified account of multiple behavioral phenomena observed in force-field experiments. 

The motor adaptation model of ref. \cite{smith06}, while being quite successful in its application to experiments, is somewhat empirical; the introduction of slow and fast modes does not have a clear microscopic basis, with the choice of the parameter values seeming arbitrary. Moreover, the linearity of the equations might be seen as a limitation on their ability to model memory. Working within the framework of activity-induced synaptic plasticity -- the standard paradigm for biological learning \cite{synplastnn00} -- an attempt is made here towards providing a theoretical basis for the work of ref. \cite{smith06}, with a description wherein the parameters are given meaning at a microscopic level. 

Activity-dependent synaptic plasticity naturally introduces some form of `competition' and `cooperation' between synapses in biological neuronal networks. These synaptic interactions occur via the patterns of activation of the pre- or post-synaptic neurons shared in common \cite{spike02}. Competition in interacting systems is generally modeled in the language of game theory \cite{game}. Here, we draw upon a game-theoretic model of competitive learning \cite{amjml99}, originally introduced in a sociological context, to propose a model for synaptic plasticity which incorporates a notion of synaptic competition. The rules of plasticity can well be viewed as `decisions' of a synapse to strengthen or weaken in response to the `outcomes' -- corresponding to activation or quiescence -- of the pair of neural units that it connects. 

For the sake of completeness, we shall briefly summarize the strategic learning model of ref.~\cite{amjml99} here, as it provides the starting point for the work presented in this article. This model was originally devised to describe social phenomena like the diffusion of innovations in connected societies; however, its essence is general enough to describe any paradigm involving decision-making in the face of competing alternatives. `Agents' are located at the sites of a regular lattice, and can be associated with one of two `types': fast (F) or slow (S); these could represent competing strategies, for example. Underpinning the dynamics of the system is the assumption that every agent revises its choice of type at regular intervals, and in this it is guided by two rules: a {\it majority} rule, reflecting the tendency of the agents to align with their local neighborhood, followed by an adaptive {\it performance-based} rule, via which the agent selects the type that it perceives to be more successful in its local neighborhood. The notion of success is gauged in terms of the random outcomes of the agents in some `game', with a favorable outcome being ascribed to every F-type (S-type) individual with an independent probability $p_+$ (resp. $p_-$). Thus, if an agent is surrounded by $N_+$ ($N_-$) nearest neighbors of the F (S) type, $I_+$ (resp. $I_-$) of which turn out to be successful in a particular trial, the agent arrives at a decision on whether or not to convert by comparing the ratios $I_+/N_+$ and $I_-/N_-$; if, for example, an agent is currently in the F state, then it will switch to the S type provided that $ I_+/N_+ < I_-/N_-$, and remain unchanged otherwise. It should be clear that the above outcome-based updates naturally introduce stochasticity into the dynamics. In ref.~\cite{amjml99}, a detailed analysis of this model was carried out under the assumption of coexistence (i.e., when $p_+ = p_- \equiv p$), and its collective behavior, as a function of the parameter $p$, was shown to exhibit multiple dynamic phases separated by critical phase transitions.   

In the following, the above ideas will be carried over to a synaptic setting. The adaptive nature of the aforementioned update rules suggests their potential applicability to modeling plastic synapses, and it will be interesting to see what consequences they might have for learning and memory, particularly in the context of making a connection with the work of ref.~\cite{smith06}.

This article proceeds as follows: In the next section, we first outline our model for synaptic plasticity. Next, we define a protocol for applying a signal to our system in its mean-field limit. We then analyze the learning and forgetting behavior in this set-up. The paper concludes with a discussion of the implications of our results, and further possible extensions.

\section*{Model and Analysis}

\begin{figure}
\begin{center}
\includegraphics[scale=1.0,height=4in,width=3.2in]{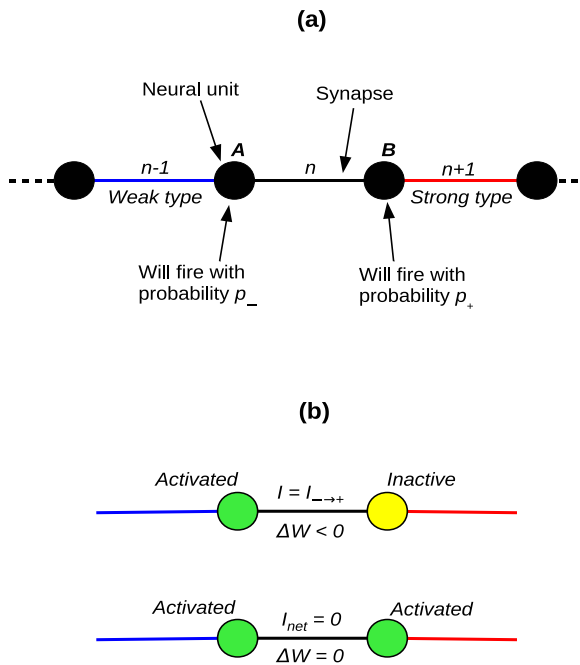}
\end{center}
\caption{\small (a) A 1-D lattice of neural units connected by undirected binary synapses. (b) Two possible `outcomes' when a synapse has a `strong' type and a `weak' type neighbor (as in (a) above). The middle synapse is under consideration (for being updated) in this case. }
\label{fig1}
\end{figure}

We consider undirected, binary synapses each of which is assumed to be of either the `strong' or the `weak' type; these synaptic states are characterized by unequal strengths (weights). Although undirected, i.e. symmetric synapses are an idealized approximation, their use has figured in several previous theoretical studies of neural networks (see e.g.~\cite{hopfield,gardner,hinton}), so this is not an unusual assumption that we are making. As for the binary nature of our synapses, this is a natural approximation to synapses possessing a finite set of discrete states, in support of which there is some experimental evidence~\cite{discrete1,discrete2}; these, again, have been used in earlier mathematical models (e.g.~\cite{fusi05,vanrossum,zecchina}). For simplicity, we choose to work with a one-dimensional chain. A pair of synapses in this set-up is treated as a pair of interacting neighbors if the synapses share a connected neural unit. The basic features of the model are sketched in fig.~\ref{fig1}. To begin with, we make a `first-order' approximation here: when a particular synapse is chosen to be considered for an update, the neural units connected to it are assumed to take on the identities of the {\it neighboring} synapses, with the effect of the synapse under consideration being omitted at this step.  In other words, the firing probability of e.g. neuron A in fig.~\ref{fig1}(a) is assumed to be determined only by the type of synapse $n-1$. A neuron receiving input via a single strong synapse is more likely to be activated than one fed by a weak synapse (input statistics being the same for both). Thus, an activation probability can be associated with any neuron depending on the type of the [single] incoming synapse: $p_+$ for a strong synapse and $p_-$ for a weak synapse. In the context of Fig. 1(a), what this ultimately means -- when it is the turn of synapse $n$ to be considered for updating -- is that if synapse $n-1$ is in the strong (weak) state, then the neuron A is going to be found active with a probability $p_+$ (resp. $p_-$), {\it irrespective} of the state of synapse $n$.\footnote{The first-order approximation made above, which essentially ignores `self-coupling' of the synapses, may be justified by means of a rough analogy with an Ising-type system of interacting spins: the local field acting on a given spin is decided only by the states of its neighbors, but the decision to flip depends on its energy, i.e. on its own sign as well. The `identities' of the connected neurons and their states of activity could be treated in analogy with a local field, with the type of the synapse under consideration being analogous to the sign of a spin.} Treated this way, the probabilities $p_\pm$, although being really a property of the synapses, come into play only through their effect on neuronal outcomes (i.e. state of activity).

%%%%%%%%%%%%%%%%%%%%%%%%%%%%%%%%%%%%%%%%%%%%%%%%%%%%%%%%%
%%%%%%%%%%%%%%%%%%%%%%%%%%%%%%%%%%%%%%%%%%%%%%%%%%%%%%%%%

We shall associate any signal (or current) transmitted by a synaptic connection with a `polarity' determined by the states as well as the types of the connected neural units. A positively (negatively) polarized signal is realized when the synapse connects a strong neuron with a weak neuron, {\it and} the strong (weak) neuron alone is activated. This event occurs with a probability $p_+ (1 - p_-)$ (resp. $p_- (1 - p_+)$), and we denote such a signal by $I_{+ \rightarrow -}$ (resp. $I_{- \rightarrow +}$). All other possible combinations of neuron types and activation states are associated with zero or unpolarized current. We now propose the following rules governing synaptic weight changes: $\Delta w > 0$ ($\Delta w < 0$) whenever there is a positively (negatively) polarized current, and $\Delta w = 0$ in all other cases. Furthermore, to be consistent with the two-state nature of the synapses, it is assumed that a strengthening event would effect a weak $\rightarrow$ strong conversion, while leaving an already strong synapse unchanged (the same logic holds for a weak synapse also). Thus, loosely speaking, the two neighboring synapses of any given synapse `compete' to decide its type, and this goes on repeatedly throughout the network.

For the sake of clarity, implementation of the above rules is illustrated with an example. Say a synapse has one neighbor of each type, as is depicted in fig.~\ref{fig1}. For this configuration, a total of four outcomes for the neuronal pair A-B are possible. There will be no weight changes if both neurons get activated (giving an unpolarized synaptic current) or if both remain silent; the likelihood of this happening is $p_+ p_- + (1 - p_+)(1 - p_-)$. A depressing event ($\Delta w < 0$) occurs if neuron A fires but neuron B remains inactive, and this has a probability $p_- (1 - p_+)$. The only remaining possibility is that neuron B gets activated and neuron A does not; this occurs with a probability $p_+ (1 - p_-)$, and is accompanied by potentiation ($\Delta w > 0$).     

Referring back to the introductory section, it is clear that the above set of rules has a close correspondence with a one-dimensional version of the outcome-based conversion rules introduced in ref. \cite{amjml99} (and this is the reason, in fact, for choosing them in the first place). At the same time, their biological reasonableness could be argued for by leaning on earlier work on rate-based models of synaptic plasticity. In such continuous-time models, the firing rate of the neuron, instead of its membrane potential, is taken as the basic dynamical variable, and synaptic plasticity is a continuous process that depends on the firing rates of the pre and post-synaptic neurons. The dynamical equation describing the time evolution of the synaptic weight usually involves some non-linear function of pre/post-synaptic activities and the weight itself, and in some cases, a dependence on averages of the firing rates over some temporal window has also been motivated~\cite{spike02,abbott,bcm}. Taking a cue from such approaches, we speculate that the rules for synaptic weight changes proposed in the previous paragraph might also be realizable through an iterative, discrete Hebbian equation symbolically expressed as $\Delta w \propto \phi ((u - \left<u\right>)(v - \left<v\right>))$, where the form of the non-linear function $\phi$ is chosen to provide an appropriate `fit' to the plasticity rules. Here, $u$ and $v$ represent the activity states of the two connected neurons, which for simplicity are assumed to be binary variables in the present set-up, being either active (1) or inactive (0). The symmetric form of the argument of $\phi$ is in keeping with the bi-directional nature of the synapses and ensures that the synaptic response is insensitive to the spatial direction of any current, while still being sensitive to its polarity. The inclusion of time averages of the activity of the connected neurons allows for a characterization of the strengths of the neighboring synapses in this picture, and hence allows for the determination of the polarity of any current at the synapse.  

The resulting model of induced synaptic plasticity is evidently stochastic,
and could be simulated by using a range of updates, as in the case
of the model of game-theoretic origin~\cite{amjml99}. 
We choose however, to tackle the problem analytically, by looking at
a mean-field version of the model. The idea behind the mean-field
approximation is that we look at the average behavior in an infinite system. 
This, at one stroke, deals with two problems: first, there are no fluctuations associated 
with system size, and second, the `first-order' approximation that we have made in ignoring the
self-coupling of the synapse, is better realized. The resulting mean-field
equations are fully deterministic in the sense that once the probabilities
$p_\pm$ are given, they admit of no further sources of stochasticity in their
solution.

In the mean-field representation, every synapse is assigned a probability (uniform over the lattice) to be either strong ($f_+$) or weak ($f_-$), so that spatial variation is ignored, as are fluctuations and correlations. This single effective degree of freedom allows for a description of the system in terms of its fixed point dynamics. The rate of change of the probability $f_+$, say (which in the limit of large system size is equivalent to the fraction of strong units), with time is computed by taking into account only the nearest-neighbor synaptic interactions, via the rules defined earlier. The dynamical equation for $f_+(t)$ assumes the following form:
\begin{eqnarray}
f_+ (t+1) &=& r_{+ \rightarrow +}(t)~ f_+ (t) ~+~ r_{- \rightarrow + }(t) ~f_- (t) \nonumber \\
                &\equiv& F (f_+ (t))
                \label{eq1}
\end{eqnarray}
with the transition probabilities being given by
\begin{eqnarray}
r_{+ \rightarrow +}(t)  &=&   f_+^2(t) + f_-^2(t) \nonumber \\
 &+& 2 f_+(t) f_-(t) \left(  1 - p_- (1 - p_+) \right) \nonumber  \\
r_{- \rightarrow + } (t) &=&   2f_+(t) f_-(t) p_+ (1 - p_-). 
\end{eqnarray}
The fractions of strong and weak types are, of course, normalized by definition: $f_+(t) + f_-(t) = 1$.

The [implicitly] time-dependent transition probabilities, which incorporate the effect of nearest-neighbor coupling, introduce non-linearity into the dynamics, an obvious departure from the linear coupled equations of \cite{smith06}. The deterministic dynamics of eq.~\ref{eq1} yields stationary states ($f_+ ^*$) to which the system would relax exponentially starting from an arbitrary initial state. Besides the `trivial' unstable fixed points at 0 and 1 corresponding to homogeneous states (all units being one or the other type), the algebraic equation $f_+^* = F(f_+ ^*)$  also possesses a stable solution; this is given by
\begin{equation}
f_+ ^* (p_+, p_-) = \frac{(1-p_-)p_+}{(p_+ + p_- - 2 p_+ p_-)}.   \label{eq3}
\end{equation}
(In the presence of fluctuations, e.g. associated with finite system sizes in
mean-field, or in the full solution of the stochastic equations, we would
expect the trivial fixed points to be absorbing, and the stable fixed point
associated with eq. \ref{eq3} to be metastable.) The time scale for relaxation to this fixed point is the other dynamically relevant quantity, which again can be extracted from eq. \ref{eq1} and is given by 
\begin{equation}
\tau = \frac{1}{2} \left [ \frac{1}{p_+ (1 - p_-)} +  \frac{1}{p_- (1 - p_+)} \right ].
\end{equation}
  
\begin{figure}[t]
\begin{center}
\includegraphics[scale=1.0,height=2.8in,width=3.1in]{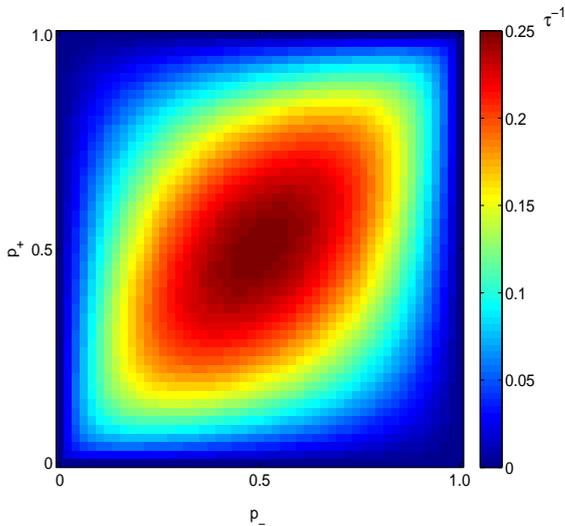}
\end{center}
\caption{\small Timescales for relaxation in the one-dimensional coarse-grained representation: a color-coded display of the variation of the [inverse of the] relaxation timescale ($\tau^{-1}$) in the ($p_+, p_-$) plane.}
\label{fig2}
\end{figure}

This relaxation time is the central quantity with regard to learning and forgetting protocols (see e.g. \cite{smith06}). It depends on the outcome probabilities $p_\pm$, and varies with the location of the corresponding fixed point. This dependence is illustrated in fig.~\ref{fig2} in the ($p_+, p_-$) plane. A possible way of defining learning and retention in the coarse-grained limit is the following: we first define a general time-dependent signal as a `perturbation' of the system parameters ($p_+, p_-$) having the following form: $(p_+, p_-) \rightarrow (p_+ + s(t), p_- - s(t))$. This choice of signal definition makes sense if one thinks of $p_+ - p_-$ as a `biasing field', in analogy with spin models, as has been suggested before \cite{amjml99}. Also, this way the signal is being applied to both the parameters, rather than preferentially to only one of them. Thus, the application of a signal of this form has the effect of introducing a time dependence into the system parameters. 

%%%%%%%%%%%%%%%%%%%%%%%%%%%%%%%%%%%%%%%%%%%%%%%%%%%%%%%%%%%%%%%%%%55
%%%%%%%%%%%%%%%%%%%%%%%%%%%%%%%%%%%%%%%%%%%%%%%%%%%%%%%%%%%

If one were to consider the case of a constant input signal, the fixed point would shift to a new location along a $p_+ + p_- =$ const. line in the ($p_+, p_-$) plane. One can then imagine a protocol whereby a constant signal is switched on at $t=0$ and persists up to a time $t=T$, following which the system reverts to its original state. Learning and forgetting are both exponential relaxation processes in this setting, and two timescales naturally enter the picture: the [learning] timescale for moving to the new stable state (after the signal is applied), and the [forgetting] timescale for reverting to the default fixed point once the signal is turned off. It may be noted that, since the relaxation timescale is a function of the parameters $p_\pm$ of the end state, it is in general different for the learning and the forgetting: the former depends on the values of $p_\pm$ in the presence of the signal, while the latter depends on the unperturbed state. 

\begin{figure}[t]
\begin{center}
\includegraphics[scale=1.0,height=2.8in,width=3.1in]{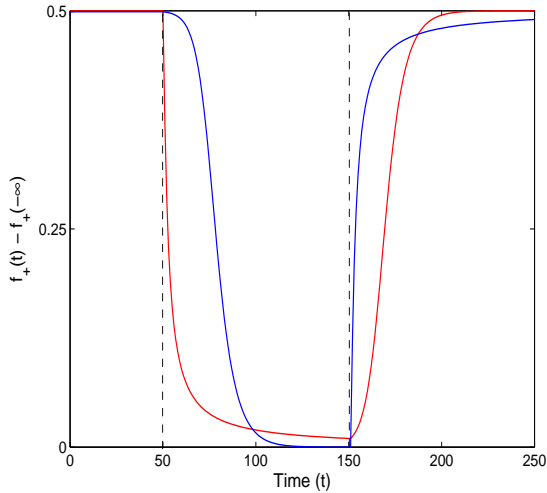}
\end{center}
\caption{\small An example of different rates for retaining/relearning a constant input signal depending on the separation of the two parameters $p_\pm$. The signal is first turned on at $t = -500$ (not shown here), turned off at $t=50$ and then re-applied at $t=150$ (indicated by dashed lines). The $y$-axis represents the variable $f_+(t)$ relative to its value at the corresponding {\it default} fixed point. The red curve corresponds to $(p_+,p_-) = (0.001, 0.999)$ and the blue curve to $(p_+,p_-) = (0.5,0.5)$. The same signal amplitude ($s = 0.499$) has been chosen in both the cases. The system that `forgets' faster `relearns' slower.}
\label{fig3}
\end{figure}

One of the strengths of the preceding approach is that performance optimization can be directly related to the microscopic parameters $p_\pm$ (in contrast to the approach of \cite{smith06} where optimization relied on the relative values of multiplicative constants). Fig.~\ref{fig2} suggests an approach to optimizing the performance, i.e. achieving long forgetting times and typically short learning times: by choosing the default parameters in such a way that the unperturbed state of the system lies near the lower right corner (or the upper left corner), the timescale for retention, being only a function of the unperturbed state, can be made very long, with the average timescale for learning applied signals being shorter. This limit corresponds to having a wide separation between the timescales associated with the two parameters $p_+^{-1}$ and $p_-^{-1}$. (If, alternatively, one were to choose the default values of $p_\pm$ to lie closer to the middle of the graph, the forgetting timescale would be shortened, an undesirable feature.) It may be noted that translating the default state along the diagonal line given by $p_+ + p_- = 1$ only modifies the retention time, while leaving the range of signals that can be absorbed (and thus the average learning time) unchanged. Additionally, given the form of the signal as defined above, which can only produce shifts parallel to the  $p_+ + p_- = 1$ diagonal, it should be clear that the range of allowed signals is maximal when the system `lives' on this diagonal, rather than on any other line parallel to it. Fig.~\ref{fig3} illustrates the system response for two example choices of the default parameters. When the parameters are well separated, the system shows more optimal behavior. 

%%%%%%%%%%%%%%%%%%%%%%%%%%%%%%%%%%%%%%%%%%%%%%%%%%%%%
 %%%%%%%%%%%%%%%%%%%%%%%%%%%%%%%%%%%%%%%%%%%%%%%%%%%%%

\section*{Discussion}

It may be recalled that the work of ref. \cite{smith06} was the inspiration for the present one, and we wished to enrich, from a physical point of view, a formalism that had an appealing link to experiment. A direct comparison of the effective equations in the two cannot be made (although they have some formal similarity). Nevertheless, the basic ingredients common to both are a notion of `memory' and two timescales, which e.g. leads to quicker relearning in the protocols explored in \cite{smith06}. We have provided a theoretical basis for the work of \cite{smith06}, by introducing a microscopic description, where the outcome-based competition between neighboring synapses leads to our mean-field representation. The nonlinearities, related to synaptic efficacies, provide a more reasonable basis for memory to be observed. Additionally, the formal similarity between our work and that of~\cite{smith06} in terms of an effective variable description suggests that our dynamical equation could also be experimentally realizable. A detailed exploration of the model presented here under different experimental protocols will be carried out in a separate study.

To summarize, we have sketched a plausible model for competing synaptic interactions mediated by neural units, in analogy with an agent-based model of adaptive strategic learning. This model provides a microscopic basis for memory; this is encapsulated in the parameters $p_\pm$ associated with the weights of the two states of a binary plastic synapse. We have, in the previous section, chosen a simple illustration of the memory manifested in our model, by devising an elementary protocol of learning and forgetting, and imposing it on our mean-field representation of synaptic dynamics. Our formalism allows us to conclude that memory is optimized if the timescales corresponding to strong and weak synapses are well separated. It is also rather general, and amenable to many possible extensions: the simplest such might be the choice of different representations of signals, or an increase in complexity for the signals and protocols. Also, one could extend the formalism to include stochasticity of synaptic conversions \cite{fusi05,amjml99}; and of course, one could choose to look at the equations beyond mean field, including for example spatial correlations. 

In conclusion, we have devised a rather general formalism for synaptic dynamics based on multiple timescales, whose results may be viewed against the backdrop of previous work underscoring the adaptive significance of having a multiplicity of timescales in biological mechanisms of learning and memory \cite{fusi05,smith06,juntani08,fusi07}.         

%%%%%%%%%%%%%%%%%%%%%%%%%%%%%%%%%%%%%%%%%%%%%%%%%%%%%%%%%
\section*{Acknowledgments}
The authors thank Dr Nicolas Brunel and Dr Jean-Marc Luck for helpful discussions. A.M. acknowledges a grant from the DST (Govt. of India) under the project ``Generativity in Cognitive Networks", through which G.M. was supported.
%%%%%%%%%%%%%%%%%%%%%%%%%%%%%%%%%%%%%%%%%%%%%%%%%%%%%%%%

\end{document}